# Topological Dirac line nodes and superconductivity coexist in SnSe at high pressure

Xuliang Chen[1,*], Pengchao Lu[2,*], Xuefei Wang[1], Yonghui Zhou[1], Chao An[1], Ying Zhou[1], Cong Xian[1], Hao Gao[2], Zhaopeng Guo[2], Changyong Park[3], Binyang Hou[3], Kunling Peng[4], Xiaoyuan Zhou[4], Yimin Xiong[1,5,†], Jian Sun[2,5,†], Zhaorong Yang[1,5,6,†], Dingyu Xing[2,5] and Yuheng Zhang[1,5]

[1]High Magnetic Field Laboratory, Chinese Academy of Sciences and University of Science and Technology of China, Hefei 230031, China

[2]National Laboratory of Solid State Microstructures, School of Physics, Nanjing University, Nanjing 210093, China

[3]HPCAT, Geophysical Laboratory, Carnegie Institution of Washington, Argonne, Illinois 60439, USA

[4]College of Physics, Chongqing University, Chongqing 401331, China

[5]Collaborative Innovation Center of Advanced Microstructures, Nanjing, 210093, China

[6]Key Laboratory of Materials Physics, Institute of Solid State Physics, Chinese Academy of Sciences, Hefei 230031, China

___________________________________________________________

*These authors contributed equally to this work.

†Correspondence should be addressed to: Y.X. (yxiong@hmfl.ac.cn) or J.S. (jiansun@nju.edu.cn) or Z.Y. (zryang@issp.ac.cn).

**Abstract**

Due to fundamental interest and potential applications in quantum computation, tremendous efforts have been invested to study topological superconductivity. However, bulk topological superconductivity seems to be difficult to realize and its mechanism is still elusive. Several possible routes to induce topological superconductivity have been proposed, including proximity efforts, doping or pressurizing a topological insulator or semimetal. Among them, the pressurizing is considered to be a "clean" way to tune the electronic structures. Here we report the discovery of a pressure-induced topological and superconducting phase of SnSe, a material which is highly focused recently due to its superior thermoelectric properties. *In situ* high-pressure electrical transport and synchrotron X-ray diffraction measurements show that the superconductivity emerges along with the formation of a *CsCl*-type structural symmetry of SnSe above around 27 GPa, with a maximum critical temperature of 3.2 K at 39 GPa. Based on *ab initio* calculations, this *CsCl*-type SnSe is predicted to be a Dirac line nodes (DLN) semimetal in the absence of spin-orbit coupling, whose DLN states are protected by the coexistence of time-reversal and inversion symmetries. These results make *CsCl*-type SnSe an interesting model platform with simple crystal symmetry to study the interplay of topological physics and superconductivity.

The search for topological superconductivity and exotic fermions such as Majorana quasiparticles has been a rapidly growing research field in condensed matter physics for both fundamental interests and potential applications in fault-tolerant topological quantum computation [1, 2]. Several methods have been proposed and applied to realize topological superconductivity, such as using the proximity effect at the interface between a topological insulator and superconductor [3, 4] or nanowire[5, 6], doping a topological insulator [7, 8], as well as compressing a topological insulator or a topological semimetal [9-17]. Among all the above methods, application of external pressure is considered as a clean and effective way to tune lattice as well as electronic states, especially for the study of topological materials and their possible bulk superconductivity [10–17]. However, the coexistence of superconductivity and topological properties under pressure is not obvious, for instance, the high pressure phase of TaAs is a new Weyl semimetal but it is not superconducting [18], and the superconductivity emerges in $WTe_2$ under pressure while its Weyl nodes disappear [19].

Like the topological insulators $Bi_2Se_3$ and $Bi_2Te_3$ [20, 21], SnSe is known as an excellent thermoelectric material, with a record figure of merit of 2.6±0.3 at 923 K along the b axis [22, 23]. Under ambient conditions, SnSe is an ordinary semiconductor and has a layered orthorhombic (*Pnma*, No. 62) *GeS*-type crystal structure [24]. Interestingly, SnSe with cubic *NaCl*-type structure was theoretically proposed to be a native topological crystalline insulator [25] with surface states protected by mirror symmetry [25–29]. By means of strain generated from lattice mismatch between SnSe and the substrate in epitaxial films, metastable *NaCl*-type SnSe was realized, displaying Dirac surface states on the basis of angle-resolved photoemission spectroscopy [30]. Under applied pressures to around 30 GPa, SnSe was reported to undergo an orthorhombic (*Pnma*) to orthorhombic (*Cmcm* or *Bbmm*, No. 63) structural transition at ~10 GPa [31–36]. Being accompanied by the structural transition, an electronic transition from a semiconducting to semimetallic state was observed [24]. However, the properties of SnSe at higher pressures remains unexplored.

Here, we show that SnSe exhibits a topological quantum phase transition and superconductivity induced by pressure. By *in situ* high pressure synchrotron X-ray diffraction and electrical transport measurements, a new phase of SnSe in *CsCl*-type structure is observed above around 27 GPa and this new phase is found to be superconducting. Theoretical calculations reveal that this *CsCl*-type SnSe has a unique electronic feature with topological Dirac line nodes, which may have contributions to the observed superconductivity, at least partially. The simultaneous presence of nontrivial topology and superconductivity makes SnSe a promising material to study the correlations between them.

## Results

**Structural phase transitions under pressure**

*In situ* angular dispersive synchrotron X-ray diffraction (XRD) experiments were performed with SnSe sample up to 50.1 GPa at room temperature ($\lambda$ = 0.4246 Å). The experimental results are presented in Fig. 1(a). Standard Rietveld refinements were performed using GSAS program [37]. Starting at 1.3 GPa, the pattern can be indexed solely by the *GeS*-type orthorhombic structure with space group *Pnma* (No. 62), which is the same as that of SnSe at ambient conditions [38]. With increasing pressure, we have identified three different phases during compression to 50.1 GPa: the low pressure Phase I remains only to 15.5 GPa (black); a pure Phase II shows up in the pressure region of 19.3-23.0 GPa (red); a broad mixture region of Phases II and III exists from 27.3 to 43.4 GPa (yellow); and above 50.1 GPa, a pure Phase III forms (blue). It should be noted that when decompressing back to 0.3 GPa (denoted by D), the XRD pattern came back to the starting structure, manifesting that the pressure-driven structural transition is reversible.

The lattice structure of Phase II can be described by space group *Bbmm* (No. 63), which is a non-standard setting of space group *Cmcm* (No. 63) in order that the crystal axes are maintained in the same orientation as for the low pressure *Pnma* structure. Such structural transition of *Pnma* → *Bbmm* (or *Cmcm*) under pressure was also reported by previous work [31-33]. In the pressure range of 27.3-43.4 GPa, the XRD pattern is a superposition of Phase II and Phase III. At 50.1 GPa, the XRD peaks for Phase III can be indexed by two structures: the *NaCl*-type with space group *Fm-3m* (No. 225) or the *CsCl*-type with space group *Pm-3m* (No. 221). Based on our following experimental analysis and theoretical calculations, however, one will see that only the latter is reasonable. Selected XRD profile refinements are presented in Fig. S1 and the corresponding fitting parameters are summarized in TABLE S1.

Figure 1(b) shows the variations in lattice parameters as a function of pressure. In Phase I, with increasing pressure all of the lattice parameters decrease gradually. However, the lattice parameter *a* is apparently more compressible than the other two, showing an anisotropic feature. This anisotropy can be related to the fact that the unit cell of SnSe is composed of bilayers perpendicular to the direction of the longest axis (a axis in our case). And the van der Waals force dominates between adjacent layers while strong chemical bond links the atoms within the bilayers. On the other hand, the lattice parameters *b* and *c* continuously approach a same value from about 6 GPa. At about 11 GPa the axial ratio changes from $b/c < 1$ to $b/c > 1$. In Phase II, while the anisotropic compressibility remains, $b/c$ keeps almost unchanged with pressure. Since the c axis is much stiffer than b axis (Fig. 1(b)), it thus is inferred that the transition from Phases I to II is related to a displacive movement of Sn/Se atoms along the [010] direction (or b axis), in accordance with previous reports [31, 34]. Note that, over the transition, all the lattice parameters evolve almost smoothly with pressure, indicating the transition to be a second order. It is also interesting to see that the lattice of SnSe approaches a tetragonal metric at around 11 GPa, but remains orthorhombic at all pressures for Phase I and Phase II. Such displacive structural transition and the modified tetragonal metric at high pressure are quite similar to those of SnSe at elevated temperature [38, 39]. Phase III starts to appear at 27.3 GPa and the lattice parameter $a$ decreases slightly with pressure up to 50.1 GPa, the highest pressure studied in the present XRD experiment. The discontinuous change of lattice parameters from phases II to III implies that the transition is a first order. A broad pressure region of coexistence between Phase II and III is observed, and thus it is hard to determine the exact critical pressure. Here the critical pressure from Phase II to III is defined as the one above which Phase III starts to appear (dashed line in Fig. 1(b)). The transition from Phase II to III involves a bonding of Sn to Se from neighboring bilayers, which makes a predominantly two-dimensional (2D) (Phase II) to a three-dimensional (3D) (Phase III) crossover.

The isothermal equations of state for Phase I, II and III are plotted in Fig. 1(c). The isothermal equations of state were fitted by the third-order Birch-Murnaghan formula [40] as indicated by the solid lines. The calculated parameters are: the bulk moduli $B_0 = 39.8$ GPa, and its first derivative $B_1 = 4.31$ for Phase I; $B_0 = 55.4$ GPa, and $B_1 = 4.55$ for Phase II; and $B_0 = 64.3$ GPa, and $B_1 = 4.81$ for Phase III. The estimated volume per unit cell at zero pressure are $V_0/Z = 53.6$ Å$^3$ for Phase I, $V_0/Z = 49.5$ Å$^3$ for Phase II, and $V_0/Z = 44.9$ Å$^3$ for Phase III. The obtained $B_0$ of Phase I is in reasonable agreement with the reported experimental and theoretical values [31, 32, 34–36]. Upon compression, the increasing bulk moduli can be related to the 2D to 3D structural transition. Meanwhile, the unit cell volume shrinks 6.9% from Phases II to III.

**Theoretical calculations**

In order to figure out these two pressure-induced structural transitions, especially the crystal structure of Phase III, we performed first principle calculations combined with crystal structure search techniques. Three energy-favorite phases were identified in our searching: Phase I (referred as α-SnSe) crystallized in the *GeS*-type structure with space group *Pnma* at pressures below 10 GPa; Phase II (β-SnSe) with space group *Bbmm* (or *Cmcm*) from 10 to 26 GPa; and the *CsCl*-type SnSe with space group *Pm-3m* above 26 GPa. The calculated enthalpy-pressure (ΔH-P) and volume-pressure (V-P) curves of these three phases are shown in Fig. 2, together with crystal schematic representations. The calculated V-P behaviors (Fig. 2(b)) are in good agreement with our experimental data (Fig. 1(c)). Here, it should be pointed out that a *NaCl*-type structure can also match exactly the positions of those four XRD peaks observed at 50.1 GPa, only if this hypothetical structure has the same Sn-Se bond length (or $\sqrt{2}$ times lattice constant) of the *CsCl*-type structure. this hypothetical *NaCl*-type structure is required to have much smaller volume/atom than β-SnSe, which should make an unusual shrinkage of lattice as high as 33.5% (6.9%, *CsCl*-type). On the other hand, based on our calculations, the volume/atom of the optimized *NaCl*-type structure near the critical pressure of 26 GPa is actually larger than the β-SnSe structure (Fig. 2(b)). Meanwhile, the *NaCl*-type phase has higher enthalpy than the *CsCl*-type under high pressure, by around 200 meV/atom at 30 GPa, shown in Fig. 2(a). In addition, phonon dispersion calculations confirm that *CsCl*-type SnSe are dynamically stable under pressure, while the *NaCl*-type structure exhibits modes with considerable negative frequencies. With all these facts, we can rule out the *NaCl*-type structure for high-pressure SnSe.

Electronic structures of *CsCl*-type SnSe are further calculated, shown in Fig. 3. Analysis of electronic structures implies that the *CsCl*-type SnSe should be a topological metal with Dirac line nodes (DLN) if spin-orbit coupling (SOC) is ignored. There are two band crossing along M-X and X-R line (Fig. 3(a)), indicating a DLN ring encircling X point on the R-X-M plane. The point group along M-X line is $C_{2v}$, which allows two different irreducible representations ($\Gamma_1$ and $\Gamma_4$) and protects these two bands from hybridization, similar for the X-R line. As there is only one irreducible representation $\Gamma_5$ belonging to the $C_{2v}$ double group, it opens a gap when SOC is concerned, as shown in Fig. 3(b). Due to the coexistence of time-reversal and inversion symmetry, one can obtain the $Z_2$ topological invariant easily by calculating parity products of occupied states at eight time-reversal-invariant momenta (TRIM) points [41, 42]. For the *CsCl*-type SnSe, it is found that the parity at X is opposite from that at all other TRIM, shown in TABLE I, attributed to the conduction $A_{2u}$ and valence $A_{1g}$ states inverted at the X. Parities at the eight TRIM ($\Gamma$, 3X, 3M, R) give the $Z_2$ topological invariant ($v_0;v_1v_2v_3$) as (1;111), revealing that there should be DLNs encircling the three inequivalent X point, which is similar to the case in $Cu_3NZn$ [41].

In general, DLNs suggest existence of topologically nontrivial surface states, or called as "flat band", which is considered as a way to induce high-temperature superconductivity [43–45]. In order to study the surface states in the *CsCl*-type SnSe, we calculate the band structures of two different slab models for the (100) surface, with 40 unit cells thick and exposing one type atom to vacuum, say Sn and Se, respectively, shown in Fig. 3(e) and (f). It can be seen that the nontrivial surface bands associated with DLN are located outside the ring enclosing $\bar{\Gamma}$. In contrast to the trivial surface states emerged from dangling bonds, these nontrivial surface states are induced by inversion and time-reversal symmetry in the bulk, thus they share the same topology in spite of different terminations. Although the *CsCl*-structured SnSe is a metal due to the fact that there are several bands crossing the Fermi level, however, from the topological point of view, there exists a continuous "band gap" and nontrivial surface states, thus this phase can be considered as a strong topological insulator-like material in the presence of SOC.

**High-pressure electronic properties of SnSe**

Our theoretical calculations have shown that high-pressure *CsCl*-type SnSe has a topological nontrivial band topology. To investigate its experimental manifestations in electronic properties, we carried out electrical transport measurements on single crystal SnSe up to 55.0 GPa. Figure 4(a-d) shows the temperature dependence of the *dc* electrical resistance of SnSe in the temperature range of 1.8-300 K. At 0.4 GPa, a semiconducting behavior (*dR/dT* < 0) is observed in the whole temperature range as that of semiconducting SnSe at ambient pressure [24, 46]. With increasing pressure, the whole resistance decreases rapidly implying a remarkable reduction in energy gap of the material with pressure (see Fig. S2). From 5.4 to 9.0 GPa while the semiconducting behavior remains at high temperatures (above ∼ 250 K), a metallic-like feature (*dR/dT* > 0) shows up at low temperatures. At 12.0 GPa, a typical metallic character is observed. At the same time, the resistance at 2 K drops about six orders of magnitude with respect to that of 0.4 GPa. These observations signal a transition of SnSe under pressure from semiconductor to metal, which is related to the crossover from *b/c* < 1 to *b/c* > 1 in the pressure range of 6.0-11.0 GPa (Fig. 1(b)). Here the critical pressure (12.0 GPa) of metallization for SnSe is in excellent accordance with the one (12.6 GPa) obtained from previous resistance experiments although the latter was performed in a temperature range of 80-270 K [24]. When the pressure goes up to 17.3 GPa, however, the resistance reveals an abnormal overall rise in addition to the metallic behavior as indicated by the dashed line in Fig. 4(b). This can be ascribed to the structural transition from Phase I to II, which is triggered by the displacive movement of Sn/Se atoms and the increase of the atomic coordination via bonding of the non-bonding in-plane Sn to Se atoms.

Upon further compression, the resistance becomes increasingly metallic. The most striking finding is that at 27.2 GPa a drop of resistance is initially observed below about 2.5 K as presented in Fig. 4(c) and in Fig. 4(d), an enlarged view of the low temperature region of resistance. This sharp drop should be a signal of a superconducting transition as zero resistance is observed at 2 K at 34.3 GPa. The critical temperature, $T_C$, is defined as the onset of the superconducting phase transition. Further increasing pressure, it first increases slightly, reaches a maximum of about 3.2 K at 39.0 GPa and then decreases gradually to 2.5 K at 55.0 GPa. It should be noted here that the pressure-induced superconductivity is only observed being accompanied by the appearance of the high pressure *CsCl*-type phase above about 27 GPa. Moreover, the magnitude of the resistance at low temperatures in the normal state keeps substantially unchanged, i.e. showing pressure- and temperature-independent behaviors (above 39.0 GPa; below ∼ 70 K) under experimental resolution.

To further confirm that the drop in resistance belongs to a superconducting transition, we measured the low temperature resistance at 39.0 GPa under applied magnetic fields parallel to the [100] direction, as displayed in Fig. 4(e). With increasing field, $T_C$ decreases gradually and the resistance drop was almost smeared out at 0.5 T, which confirms that the drop of resistance is the superconducting transition. In addition, the resistance versus temperature curves are almost parallel to each other, suggesting that the flux creep effects can be ignored in the vortex dynamics of superconductivity. We have plotted the upper critical field $H_{C2}$ as a function of temperature in Fig. 4(f). Here $H_{C2}(T)$ is defined from the resistance criterion of $R_{cri} = 90\%R_n$ ($R_n$ is the normal state resistance near $T_C$). One can see that $H_{C2}$ follows a linear dependence in temperature (red dashed line). By extrapolating it to zero temperature, $H_{C2}(0)$ is estimated to be 1.12 T. Here, the Pauli paramagnetic effect does not apply since the Pauli limiting field is about 5.4 T (1.84 × $T_C$), suggestive of an absence of Pauli pair breaking.

## Discussion

As outlined in Fig. 4(f), the present XRD results evidence a set of structural transitions upon compression from the starting *Pnma* to *Bbmm* (or equivalently *Cmcm*), and over a very broad pressure range further to a *CsCl*-type structure (*Pm-3m*). The transition of *Pnma* → *Bbmm* mainly refers to a displacive movement of Sn/Se atoms along the [010] direction (or b axis) while *Bbmm* → *Pm-3m* involves additionally Sn-Se bonding between adjacent layers, owing to anisotropic compression. Correspondingly, the nearest coordination numbers of Sn increase sequentially from three to five, and five to eight (Fig. 2(c)). This makes the transition of SnSe from a predominant 2D material to a 3D network, which can also be manifested in the increasing bulk moduli under compression (Fig. 1(c)). Furthermore, the observed structural phase transitions are consistent well with the variation of resistance with pressure: in Phase I, the resistance of SnSe decreases monotonously upon compression, with a rapid closure of its band gap (Fig. S2) followed by metallization (Fig. 4(a)); when SnSe entering into the *Bbmm* modification (Phase II), the magnitude of the resistance shows an abnormal upshift (Fig. 4(b)); most strikingly, superconductivity is observed being accompanied by the appearance of the *CsCl*-type structure (Phase III). It should be noted that with increasing pressure, superconductivity first shows up at 27.2 GPa and zero resistance can only be observed above 34.3 GPa (Fig. 4(d)), which can be related to the coexistence of the non-superconducting *Bbmm* and superconducting *Pm-3m* structures.

Recently, a large amount of effort has been devoted to exploring possible topological superconductivity from topological materials by external pressure, such as $Bi_2Se_3$[12, 47], $Bi_2Te_3$[10, 48], $ZrTe_5$[16], $Cd_3As_2$[17], Sr doped $Bi_2Se_3$[49], etc. However, one may find that the appearance of superconductivity is generally along with a structural transition, making it an open question on the topology of the pressure-induced superconducting states in these materials. Here, the coexistence of the nontrivial topological feature of electronic band and superconductivity should be of special interest. In addition, the resistance in the normal state reveals temperature (below ∼70 K) and pressure (above 39.0 GPa) independent behaviors under experimental resolution. As a matter of fact, such temperature independent electronic transport behavior has also been observed in other topological materials, which was related to the presence of topological surface states [50, 51].

Finally, we would like to mention that the high-pressure modification of *Bbmm* SnSe resembles closely that of SnSe at elevated temperature [39], with an intersection of axes and approaching a tetragonal metric. Upon further warming, unfortunately, a sublimation happens before a possible *CsCl*-type SnSe may generate. As a result, the present study may offer a promising strategy for synthesizing *CsCl*-type SnSe via a combination of high pressure and high temperature techniques, and thereby study the topological nature of electronic states and their correlations to superconductivity.

In summary, we observed the formation of a new topological and superconducting *CsCl*-type phase of SnSe under high pressures based on experiments coupled with theoretical calculations. We found that this new phase appears at 27 GPa and coexists with the *Bbmm* phase in a broad pressure range, and eventually a pure *CsCl*-type phase establishes above 50 GPa. Our electronic transport measurements discovered a pressure-driven superconductivity rooted in this *CsCl*-type phase. Through *ab initio* electronic structure calculations, we predict that the $Z_2$ topological invariant of this *CsCl*-type SnSe is odd and Dirac line nodes can exist in this phase when SOC is ignored. These Dirac line nodes introduce nontrivial topological surface states, which may have contributions to the enhancement of superconductivity. It is thus proposed that SnSe may provide a model platform with high crystal symmetry to investigate the topological characteristic of electronic structures and its correlation to superconductivity.

# Methods

**Single-crystal growth**

High-quality single crystals of SnSe were grown by chemical vapor transport. Firstly, the polycrystalline powders were synthesized by a solid-state reaction using elemental Sn (Alfa Aesar 99.8%) and Se (Alfa Aesar 99.5%) in sealed and evacuated quartz ampoules. The stoichiometric amounts of mixture were heated to 700°C in rate of 3 K/min and maintained at this temperature for 3 days. Then the SnSe powder was sealed in a quartz tube under vacuum and kept in a temperature gradient 700 to 800°C for 7 days. SnSe single crystals were obtained at the hot end. The laboratory X-ray diffraction measurements, which were done at room temperature using Cu $K_\alpha$ radiation on Rigaku TTR3 diffractometer have proven that the obtained crystals are single phase with the orthorhombic structure of space group *Pnma*.

*Ab initio* **calculations**

We used *ab initio* random structure searching (AIRSS) method [52, 53] to find energy-favorite structures of SnSe. Crystal optimization and electronic structure calculations were performed using projector augmented wave (PAW) with GGA-PBE functional [54] implemented in the Vienna ab initio simulation package (VASP) [55]. The plane wave cutoff was set to 400 eV, and the Brillouin zones (BZ) were sampled by Monkhorst-Pack method with a spacing of 0.03×2π Å⁻¹. Structural optimization was performed without spin-orbit coupling, until the Hellman Feynman force were within 0.005 eV/Å. Slab models are built with 40 unit cells and a vacuum layer of 20 Å.

**High Pressure experiments**

High-pressure resistance experiments were conducted in a diamond anvil cell made of Pe-Cu alloy [16] using sodium chloride (NaCl) powder as the pressure-transmitting medium. SnSe single crystal was cleaved along the [100] direction then a tiny sheet of 100μm×50μm×10μm was loaded together with a ruby ball. Standard four-probe method was employed to measure the resistance.

High-pressure synchrotron X-ray diffraction measurements were performed in a Mao-Bell cell at 16 BM-D, HPCAT [56] at Advanced Photon Source of Argonne National Laboratory. The X-ray wavelength is 0.4246 Å and Daphne 7373 was used as the pressure medium. A two dimensional area detector Mar345 was used to collect the powder diffraction patterns. The Dioptas [57] and GSAS+EXPGUI programs [37] were employed for image integrations and XRD profile Rietveld refinements, respectively. Pressure was calibrated by using the ruby fluorescence shift at room temperature [58].

**ACKNOWLEDGMENTS**

The authors thank Prof. Xianhui Chen for stimulating discussions. This work is supported by the Ministry of Science and Technology of China (National Basic Research Program Nos. 2016YFA0300404, 2016YFA0401804 and 2015CB921202), and the National Natural Science Foundation of China (Grant Nos. 11574323, 11474288, 51472036, 51372112, 11574133, U1332143), NSF Jiangsu Province (BK20150012), Special Program for Applied Research on Super Computation of the NSFC-Guangdong Joint Fund (the 2nd phase). Part of the calculations were performed on the supercomputer at HPCC in Nanjing university and "Tianhe 2" at NSCC Guangzhou. Y.X. thanks the support of the Hundred Talents Program of the Chinese Academy of Science. The X-ray work was performed at HPCAT (Sector 16), Advanced Photon Source (APS), Argonne National Laboratory. HPCAT operations are supported by DOE-NNSA under Award No. DE-NA0001974 and DOE-BES under Award No. DE-FG02-99ER45775, with partial instrumentation funding by NSF. The Advanced Photon Source is a U.S. Department of Energy (DOE) Office of Science User Facility operated for the DOE Office of Science by Argonne National Laboratory under Contract No. DE-AC02-06CH11357.

# References

[1] Qi, X.-L. & Zhang S.-C. Topological insulators and superconductors. *Rev. Mod. Phys.* **83**, 1057 (2011).

[2] Hasan, M. Z., Xu, S.-Y. & Bian, G. Topological insulators, topological superconductors and Weyl fermion semimetals: discoveries, perspectives and outlooks. *Phys. Scr.* **T168**, 019501 (2016).

[3] Fu, L. & Kane, C. L. Superconducting Proximity Effect and Majorana Fermions at the Surface of a Topological Insulator. *Phys. Rev. Lett.* **100**, 096407 (2008).

[4] Wang, M.-X. *et al.* The Coexistence of Superconductivity and Topological Order in the $Bi_2Se_3$ Thin Films. *Science* **336**, 52-55 (2012).

[5] Mourik, V. *et al.* Signatures of Majorana Fermions in Hybrid Superconductor-Semiconductor Nanowire Devices. *Science* **336**, 1003-1007 (2012).

[6] Das, A. *et al.* Zero-bias peaks and splitting in an Al–InAs nanowire topological superconductor as a signature of Majorana fermions. *Nat. Phys.* **8**, 887 (2012).

[7] Sasaki, S. *et al.* Topological Superconductivity in $Cu_xBi_2Se_3$. *Phys. Rev. Lett.* **107**, 217001 (2011).

[8] Wray, L. A. *et al.* Observation of topological order in a superconducting doped topological insulator. *Nat. Phys.* **6**, 855 (2010).

[9] Bay, T. V. *et al.* Superconductivity in the Doped Topological Insulator $Cu_xBi_2Se_3$ under High Pressure. *Phys. Rev. Lett.* **108**, 057001 (2012).

[10] Zhang, J. L. *et al.* Pressure-induced superconductivity in topological parent compound $Bi_2Te_3$.

*Proc. Natl. Acad. Sci. U.S.A.* **108**, 24 (2011).

[11] Zhang, C. *et al.* Phase diagram of a pressure-induced superconducting state and its relation to the Hall coefficient of $Bi_2Te_3$ single crystals. *Phys. Rev. B* **83**, 140504(R) (2011).

[12] Kirshenbaum, K. *et al.* Pressure-induced unconventional superconducting phase in the topological insulator $Bi_2Se_3$. *Phys. Rev. Lett.* **111**, 087001 (2013).

[13] Kang, D. F. *et al.* Superconductivity emerging from a suppressed large magnetoresistant state in tungsten ditelluride. *Nat. Commun.* **6**, 7804 (2015).

[14] Pan, X. C. *et al.* Pressure-driven dome-shaped superconductivity and electronic structural evolution in tungsten ditelluride. *Nat. Commun.* **6**, 7805 (2015).

[15] Qi, Y. P. *et al.* Superconductivity in Weyl semimetal candidate $MoTe_2$. *Nat. Commun.* **7**, 11038 (2016).

[16] Zhou, Y. H. *et al.* Pressure-induced superconductivity in a three-dimensional topological material $ZrTe_5$. *Proc. Natl. Acad. Sci. USA* **113**, 2904-2909 (2016).

[17] He, L. P. *et al.* Pressure-induced superconductivity in the three-dimensional topological Dirac semimetal $Cd_3As_2$. *npj Quantum Materials* **1**, 16014 (2016).

[18] Zhou, Y. H. *et al.* Pressure-Induced New Topological Weyl Semimetal Phase in TaAs. *Phys. Rev. Lett.* **117**, 146402 (2016).

[19] Lu, P. C. *et al.* Origin of superconductivity in the Weyl semimetal $WTe_2$ under pressure. *Phys. Rev. B* **94**, 224512 (2016).

[20] Ghaemi, P., Mong, R. S. K. & Moore, J. E. In-plane transport and enhanced thermoelectric

performance in thin films of the topological insulators $Bi_2Te_3$ and $Bi_2Se_3$. *Phys. Rev. Lett.* **105**, 166603 (2010).

[21] Hinsche, N. F., Yavorsky, B. Y., Mertig, I. & Zahn, P. Influence of strain on anisotropic thermoelectric transport in $Bi_2Te_3$ and $Sb_2Te_3$. *Phys. Rev. B* **84**, 165214 (2011).

[22] Zhao, L. D. *et al.* Ultralow thermal conductivity and high thermoelectric figure of merit in SnSe crystals. *Nature* **508**, 373-377 (2014).

[23] Zhao, L. D. *et al.* Ultrahigh power factor and thermoelectric performance in hole-doped single-crystal SnSe. *Science* **351**, 141-144 (2016).

[24] Yan, J. J. *et al.* Pressure-driven semiconducting-semimetallic transition in SnSe. *Phys. Chem. Chem. Phys.* **18**, 5012-5018 (2016).

[25] Sun, Y. *et al.* Rocksalt SnS and SnSe: Native topological crystalline insulators. *Phys. Rev. B* **88**, 235122 (2013).

[26] Neupane, M. *et al.* Topological phase diagram and saddle point singularity in a tunable topological crystalline insulator. *Phys. Rev. B* **92**, 075131 (2015).

[27] Fu, L. Topological crystalline insulators. *Phys. Rev. Lett.* **106**, 106802 (2011).

[28] Ando, Y. & Fu, L. Topological Crystalline Insulators and Topological Superconductors: From Concepts to Materials. *Annu. Rev. Condens. Matter Phys.* **6**, 361-381 (2015).

[29] Hsieh, T. H. *et al.* Topological crystalline insulators in the SnTe material class. *Nat. Commun.* **3**, 982 (2012).

[30] Wang, Z. Y. *et al.* Molecular Beam Epitaxy-Grown SnSe in the Rock-Salt Structure: An Artificial Topological Crystalline Insulator Material. *Adv. Mater.* **27**, 4150-4154 (2015).

[31] Chattopadhyay, T., Werner, A. & von Schnering, H. G. Temperature and pressure induced phase transition in IV-VI compounds. *Revue Phys. Appl.* **19**, 807-813 (1984).

[32] Loa, I. *et al.* Structural changes in thermoelectric SnSe at high pressures. *J. Phys.: Condens. Matter* **27**, 072202 (2015).

[33] Zhang, J. *et al.* Plasma-assisted synthesis and pressure-induced structural transition of single-crystalline SnSe nanosheets. *Nanoscale* **7**, 10807-10816 (2015).

[34] de Souza, S. M. *et al.* Pressure-induced polymorphism in nanostructured SnSe. *J. Appl. Cryst.* **49**, 213-221 (2016).

[35] Alptekin, S. Structural phase transition of SnSe under uniaxial stress and hydrostatic pressure: an *ab initio* study. *J. Mol. Model* **17**, 2989-2994 (2011).

[36] Makinistian, L. & Albanesi, E. A. Study of the hydrostatic pressure on orthorhombic IV-VI compounds including many-body effects. *Computational Materials Science* **50**, 2872-2879 (2011).

[37] Von Dreele, R. B. & Larson, A. C. General structure analysis system. *Regents of the University of California* (2001); Toby, B. H. EXPGUI, a graphical user interface for GSAS. *J. Appl. Cryst.* **34**, 210-213 (2001).

[38] Chattopadhyay, T. *et al.* Neutron diffraction study of the structural phase transition in SnS and SnSe. *J. Phys. Chem. Solids* **47**, 879-885 (1986).

[39] Wiedemeier, H. & Csillag, F. J. The thermal expansion and high temperature transformation of SnS and SnSe. *Zeitschrift für Kristallographie* **149**, 17-29 (1979).

[40] Birch, F. Finite Elastic Strain of Cubic Crystals. *Phys. Rev.* **71**, 809-824 (1947).

[41] Kim, Y. *et al.* Dirac Line Nodes in Inversion-Symmetric Crystals. *Phys. Rev. Lett.* **115**, 036806 (2015).

[42] Yu, R. *et al.* Topological Node-Line Semimetal and Dirac Semimetal State in Antiperovskite $Cu_3PdN$. *Phys. Rev. Lett.* **115**, 036807 (2015).

[43] Kopnin, N. B., Heikkilä, T. T. & Volovik, G. E. High-temperature surface superconductivity in topological flat-band systems. *Phys. Rev. B* **83**, 220503(R) (2011).

[44] Volovik, G. E. From standard model of particle physics to room-temperature superconductivity. *Phys. Scr.* **T164**, 014014 (2015).

[45] Heikkilä, T. T. & Volovik, G. E. Flat bands as a route to high-temperature superconductivity in graphite. Preprint at arXiv:1504.05824.

[46] Nassary, M. M. The electrical conduction mechanisms and thermoelectric power of SnSe single crystals. *Turk. J. Phys.* **33**, 201-208 (2009).

[47] Yu, Z. H. *et al.* Structural phase transitions in $Bi_2Se_3$ under high pressure. *Sci. Rep.* **5**, 15939 (2015).

[48] Matsubayashi, K. *et al.* Superconductivity in the topological insulator $Bi_2Te_3$ under hydrostatic pressure. *Phys. Rev. B* **90**, 125126 (2014).

[49] Zhou, Y. H. *et al.* Pressure-induced reemergence of superconductivity in topological insulator $Sr_{0.065}Bi_2Se_3$. *Phys. Rev. B* **93**, 144514 (2016).

[50] Gabáni, S. *et al.* Pressure-induced Fermi-liquid behavior in the Kondo insulator $SmB_6$: Possible transition through a quantum critical point. *Phys. Rev. B* **67**, 172406 (2013).

[51] Tafti, F. F. *et al.* Resistivity plateau and extreme magnetoresistance in LaSb. *Nat. Phys.* **12**,

272-277 (2015).

[52] Pickard, C. J. & Needs, R. J. High-Pressure Phases of Silane. *Phys. Rev. Lett.* **97**, 045504 (2006).

[53] Pickard, C. J. & Needs, R. J. Ab initio random structure searching. *J. Phys. Condens. Mat.* **23**, 053201 (2011).

[54] Perdew, J. P., Burke, K. & Ernzerhof, M. Generalized Gradient Approximation Made Simple. *Phys. Rev. Lett.* **77**, 3865 (1996).

[55] Kresse, G. & Furthmüller, J. Efficient iterative schemes for *ab initio* total-energy calculations using a plane-wave basis set. *Phys. Rev. B* **54**, 11169 (1996).

[56] Park, C. Y. *et al.* New developments in micro-X-ray diffraction and X-ray absorption spectroscopy for high-pressure research at 16-BM-D at the Advanced Photon Source. *Rev. Sci. Instrum.* **86**, 072205 (2015).

[57] Prescher, C. & Prakapenka, V. B. DIOPTAS: a program for reduction of two-dimensional X-ray diffraction data and data exploration. *High Pressure Research* **35**, 223-230 (2015).

[58] Mao, H. K., Xu, J. & Bell, P. M. Calibration of the ruby pressure gauge to 800 kbar under quasi-hydrostatic conditions. *J. Geophys. Res.* **91**, 4673-4676 (1986).

272-277 (2015).

[52] Pickard, C. J. & Needs, R. J. High-Pressure Phases of Silane. *Phys. Rev. Lett.* **97**, 045504 (2006).

[53] Pickard, C. J. & Needs, R. J. Ab initio random structure searching. *J. Phys. Condens. Mat.* **23**, 053201 (2011).

[54] Perdew, J. P., Burke, K. & Ernzerhof, M. Generalized Gradient Approximation Made Simple. *Phys. Rev. Lett.* **77**, 3865 (1996).

[55] Kresse, G. & Furthmüller, J. Efficient iterative schemes for *ab initio* total-energy calculations using a plane-wave basis set. *Phys. Rev. B* **54**, 11169 (1996).

[56] Park, C. Y. *et al.* New developments in micro-X-ray diffraction and X-ray absorption spectroscopy for high-pressure research at 16-BM-D at the Advanced Photon Source. *Rev. Sci. Instrum.* **86**, 072205 (2015).

[57] Prescher, C. & Prakapenka, V. B. DIOPTAS: a program for reduction of two-dimensional X-ray diffraction data and data exploration. *High Pressure Research* **35**, 223-230 (2015).

[58] Mao, H. K., Xu, J. & Bell, P. M. Calibration of the ruby pressure gauge to 800 kbar under quasi-hydrostatic conditions. *J. Geophys. Res.* **91**, 4673-4676 (1986).

**Figure Captions**

**Figure 1 | Structural information of SnSe under pressure.** (a) The patterns were recorded at room temperature with a wavelength of λ = 0.4246 Å. Upon compression, the patterns can be well indexed by space groups *Pnma* (No. 62, below 15.5 GPa, Phase I), *Bbmm* (No. 63, between 19.3 and 23.0 GPa, Phase II) and *Pm-3m* (No. 221, above 50.1 GPa, Phase III). In the pressure range of 27.3-43.4 GPa, the XRD patterns contains a mixture of Phase II and III. When decompressing back to 0.3 GPa (denoted by D), the XRD pattern came back to the starting structure, manifesting that the pressure-driven structural transition is reversible. (b) Lattice parameters as a function of pressure. The continuous and dis-continuous changes at the low and high critical pressures (vertical dashed lines) indicate that the transitions are of second and first orders in nature, respectively. (c) The compression data was fitted by the third order Birch-Murnaghan equation of state (solid lines).

**Figure 2 | Theoretical random structure searching for SnSe under pressure.** (a) Calculated enthalpy relative to that of the β-SnSe (*Bbmm*) at pressure up to 50 GPa. (b) Unit cell volume versus pressure. (c) Crystal structures of α-SnSe (*Pnma*), β-SnSe and *CsCl*-type SnSe (*Pm-3m*). With increasing pressure, there are two transitions from α-SnSe to β-SnSe at around 10 GPa and further to *CsCl*-type SnSe at about 26 GPa.

**Figure 3 | Calculated electronic structures of bulk *CsCl*-type SnSe at 40 GPa** (a) without SOC and (b) with SOC. (c) Bulk and projected (001) surface Brillouin zone (BZ). (d) Parity products of occupied states at the eight TRIM points in the first octant of the BZ. Surface band structures without SOC for the (100) surface along high-symmetry lines, with Sn (e) and Se (f) terminations, respectively. The surface states are denoted by red curves. The nontrivial

topological surface states are indicated by black arrows. They share the same topology regardless of terminations.

**Figure 4 | Experimental evidence for pressure induced superconductivity.** (a-c) Temperature dependent resistance for SnSe single crystal. (d) An enlarged view of the low temperature resistance above 27.2 GPa, showing the superconducting transition. Note that the magnitude of the resistance at 17.3 GPa enhances abruptly in the whole temperature region as indicated by the dashed line in (b), and then decreases gradually followed by an almost unchangeable behavior above 39.0 GPa. (e) The superconducting transition of the SnSe single crystal in magnetic fields up to 0.5 T. The applied magnetic field is parallel to the [100] direction of SnSe single crystal. (f) Temperature dependence of the upper critical field $H_{C2}$. $T_C$ is determined as the 90% drop of the normal state resistance. The dashed line is a linear fit to the data and $H_{C2}(0)$ is estimated to be 1.12 T. (g) Temperature versus pressure phase diagram. The left axis stands for the resistance $R$ at 2 and 300 K and the right axis corresponds to temperature $T$.

TABLE I: Parity product of occupied states at the TRIM points, indicating the $Z_2$ indices as (1;111).

| TRIM | $\Gamma$ | R | M (×3) | X (×3) |
|---|---|---|---|---|
| Parity product | - | - | - | + |

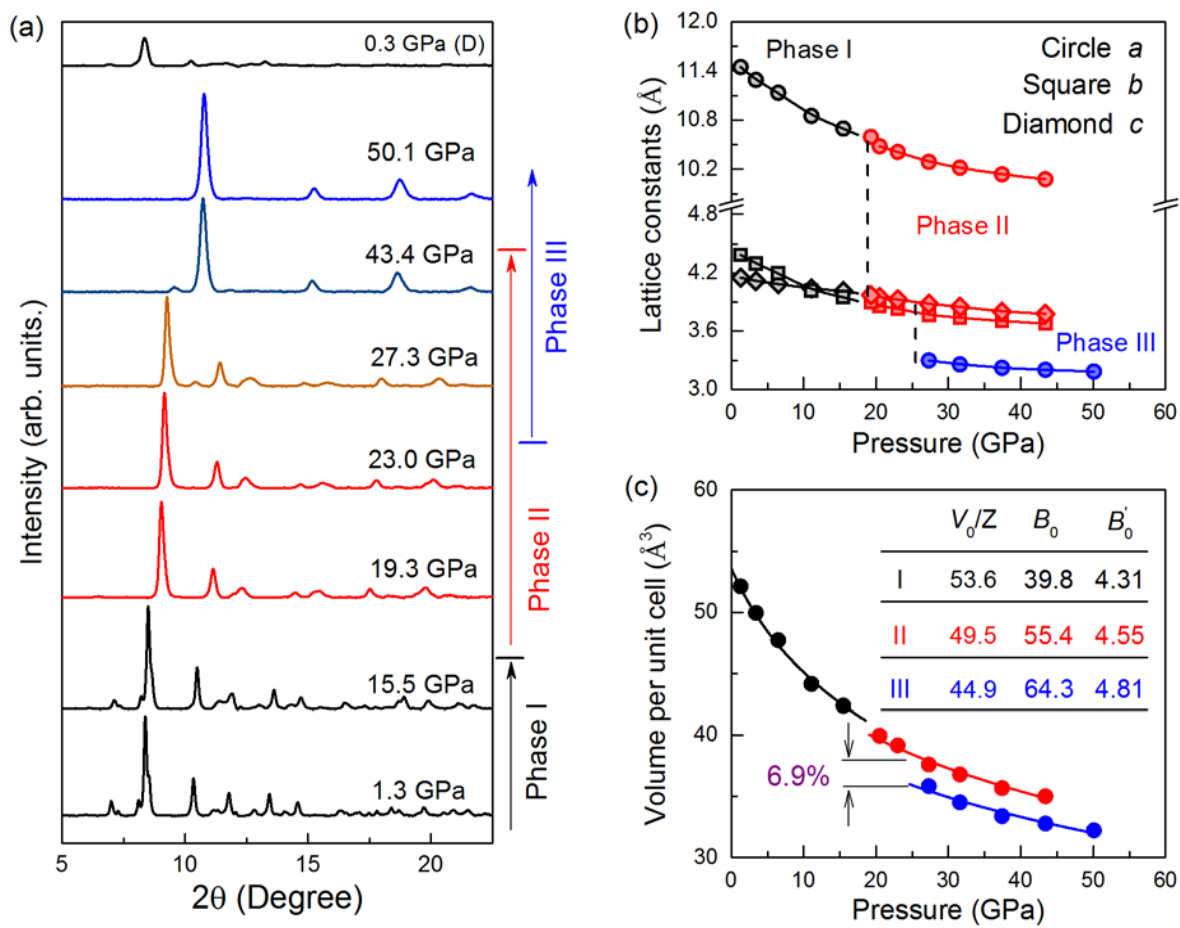

Fig. 1 Chen *et al*.

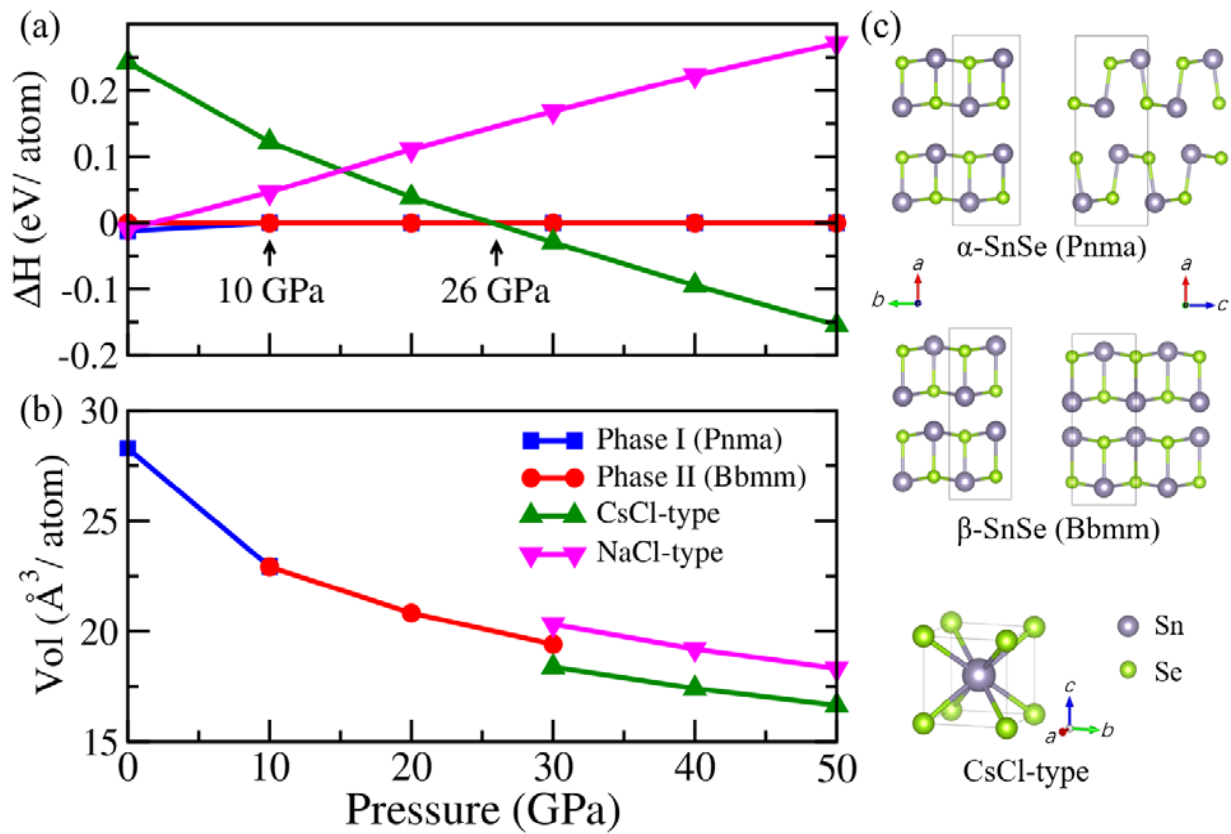

Fig. 2 Chen *et al.*

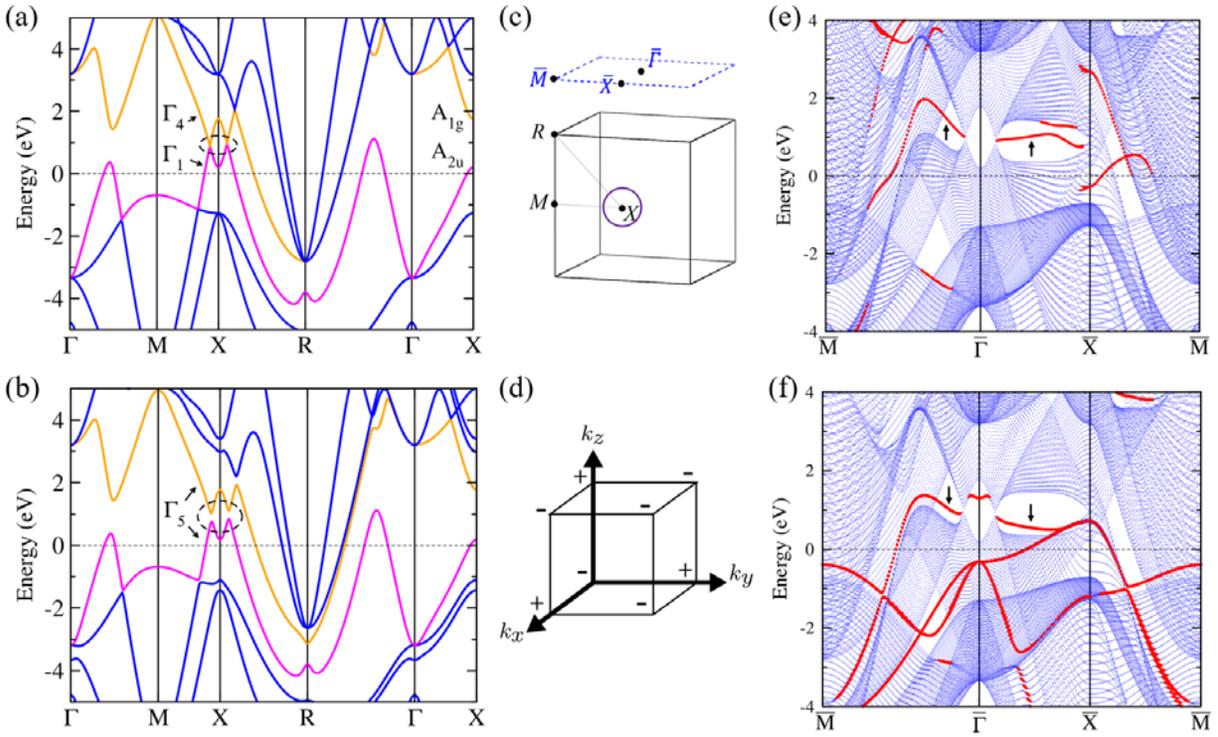

Fig. 3 Chen *et al.*

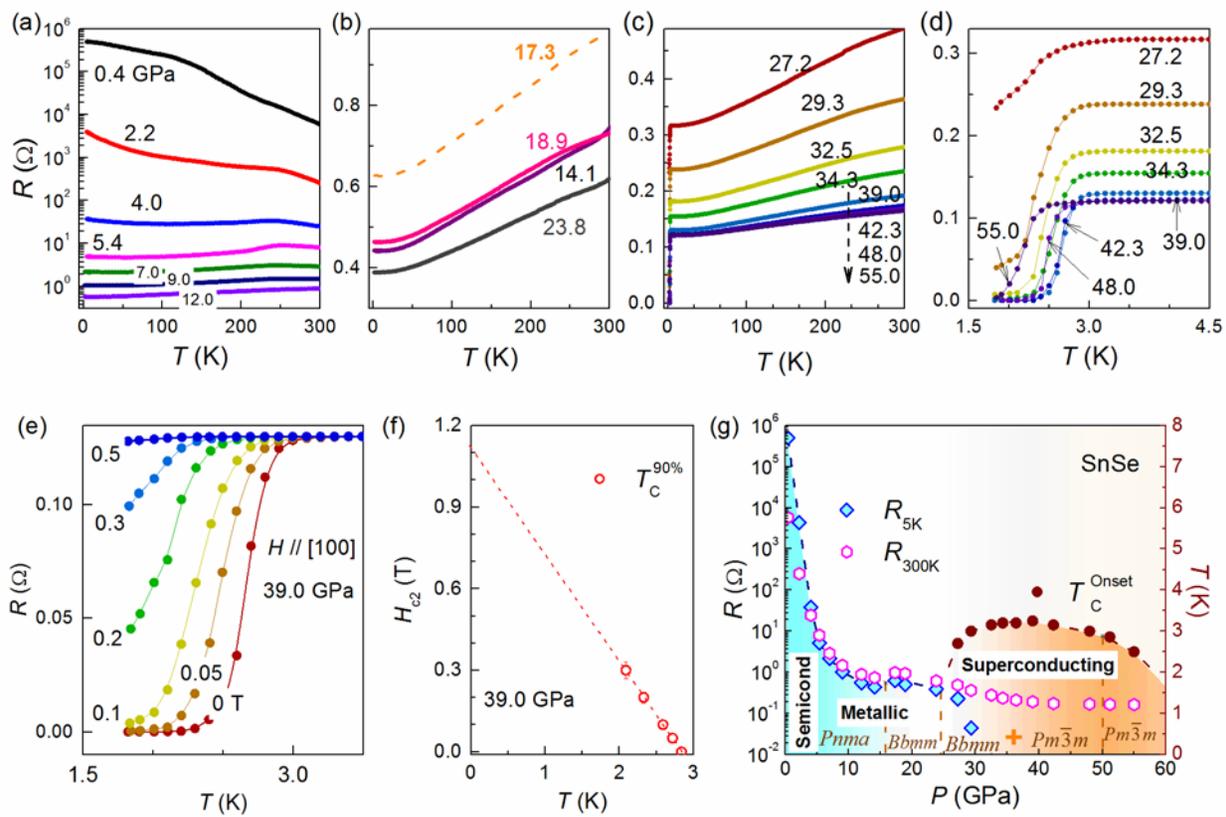

Fig. 4 Chen *et al*.

# Supplementary Figures and Tables

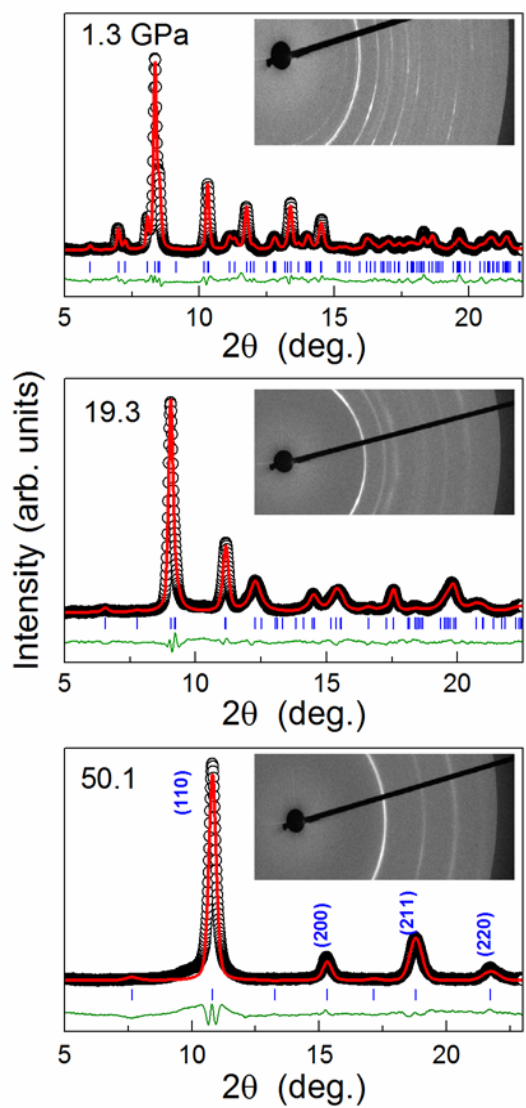

**Supplementary Figure S1. Fits of X-ray diffraction profiles.** Standard Rietveld method was used to fit the data and the refinements were performed using the GSAS program. Representative experimental raw images (partial) for each phase are also displayed.

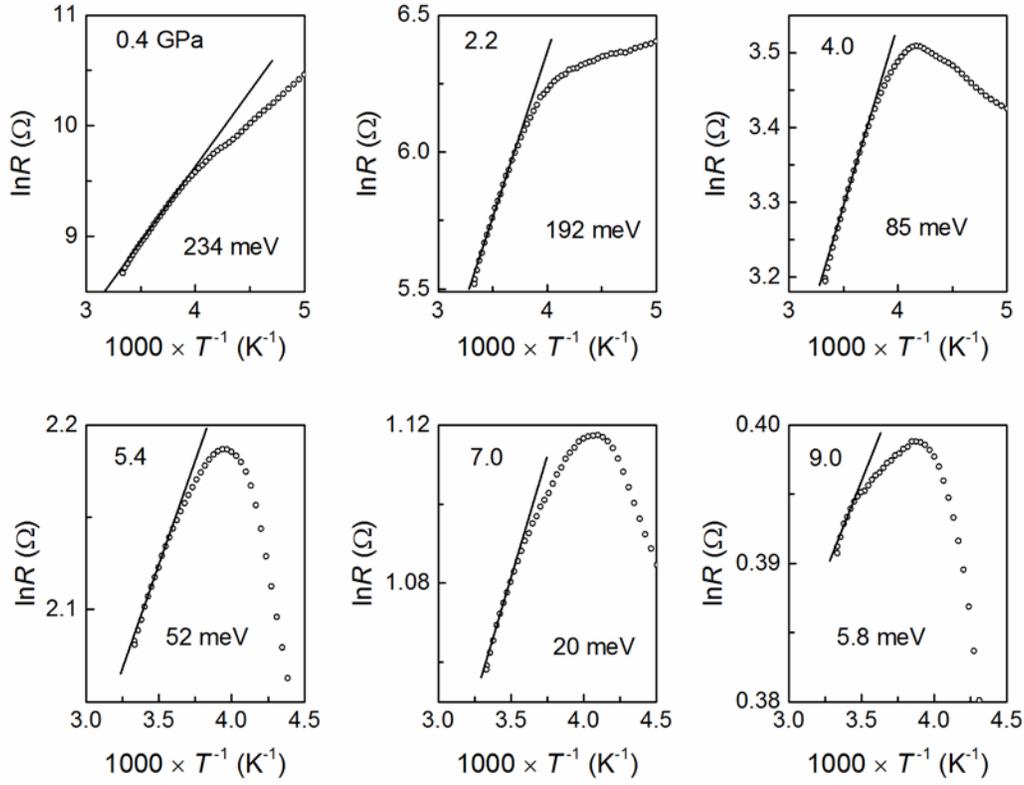

**Supplementary Figure S2. Fitting of energy gap under pressure.** The data is displayed in the representative of *lnR* versus 1000/T. The measured *dc* resistance $R$ is fitted by the well-known Arrhenius relationship of $R = R_0 \times \exp(\Delta E/k_B T)$, where $R_0$ is a pre-exponential factor and $\Delta E$ represents the activation energy, $k_B$ is the Boltzmann constant and $T$ is the absolute temperature.

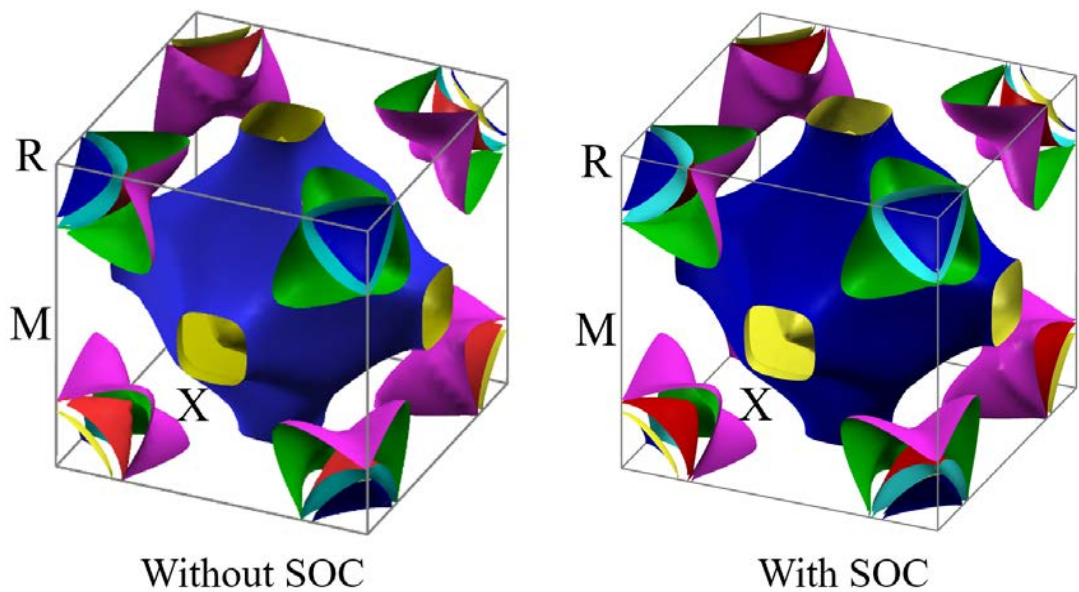

**Supplementary Figure S3.** Calculated Fermi surface of *CsCl*-type SnSe at 40 GPa without SOC (left) and with SOC (right).

**Supplementary TABLE S1:** Results of the Rietveld refinements of SnSe at 1.3, 19.3, and 50.1 GPa: lattice parameters *a-c*; the number of chemical formulas contained in the unit cell *Z*; fractional coordinates *x*, *y* and *z* of the Sn and Se positions; isotropic atomic displacement parameters $U_{iso}$; refined values $R_{WP}$, $R_P$, $\chi^2$.

| Parameters | Phase I: *Pnma* 1.3 GPa | Phase II: *Bbmm* 19.3 GPa | Phase III: Pm-3m 50.1 GPa |
|---|---|---|---|
| *a* (Å) | 11.4473(9) | 10.5928(2) | 3.1821(4) |
| *b* (Å) | 4.1517(1) | 3.9003(3) | 3.1821(4) |
| *c* (Å) | 4.3865(2) | 3.9716(6) | 3.1821(4) |
| Z | 4 | 4 | 1 |
| *x*(Sn/Se) (f.c.) | 0.1214(3)/0.8575(5) | 0.1331(5)/0.8983(12) | 0.5/0 |
| *y*(Sn/Se) (f.c.) | 0.25/0.25 | 0.25/0.25 | 0.5/0 |
| *z*(Sn/Se) (f.c.) | 0.9095(7)/0.5176(8) | 0/0.5 | 0.5/0 |
| $U_{iso}$ (Å$^2$) | 0.032(0)/0.01 | 0.0362(0)/0.01 | 0.0263(1)/0.01 |
| $R_{WP}$, $R_P$, $\chi^2$ | 0.77%, 0.6%, 0.044 | 0.66%, 0.49%, 0.031 | 1.42%, 0.99%, 0.134 |

*Note*: The atomic displacement parameter $U_{iso}$ (Se) was fixed at 0.01 Å$^2$ otherwise refine to negative values.